\documentclass{birkjour_t2}
\usepackage{draftwatermark}
\usepackage{pdfsync}
\usepackage{amsmath,amsfonts,amssymb}
\usepackage{stmaryrd}
\usepackage{graphicx}
\usepackage{verbatim}
\usepackage{color}

\def \tr{\mbox{tr\hskip 1pt}}
\def \grad{\mbox{grad\hskip 1pt}}
\def \Grad{\mbox{Grad\hskip 1pt}}
\def \det{\mbox{det\hskip 1pt}}

\begin{document}
\title[On the extended use of Backus average]{On the extended use of Backus average}
\author{Andrey Melnikov}
\address{Department of Earth Sciences, Memorial University of Newfoundland,  Canada;
{\tt amelnikov@mun.ca}}
\SetWatermarkText{MANUSCRIPT}
\SetWatermarkScale{4}

\begin{abstract}
Backus (1962)~\cite{backus} developed his technique for homogenization of a layered structure solely within the context of linear elastic theory.
In this paper we propose an extended use of Backus average~\cite{backus} for finitely deformed materials of a layered structure. We attempt to use two different approaches to account for large deformations. The first approach utilizes the connections between linear and nonlinear transverse elasticity. For the second approach we use a formulation based on prestress in the material. We conclude that the first approach, although with some limitations, can be used successfully.
\end{abstract}
%%%%%%%%%%%%%%%%%%%%%%%%%%%%%%%%%%%
\maketitle
\section{Introduction}
In linear elasticity the overall properties of a layered elastic continuum can be found using formulas described in Backus~\cite{backus}.
Each individual layer is isotropic and thus characterised by the two elasticity parameters $c_{1111}$ and $c_{2323}$.
According to Backus~\cite{backus}, the overall averaged continuum will have a transverse isotropic symmetry and thus can be described by the five elastic constants
\begin{equation}
\label{eq:elast-const1}
c^{\overline{\rm
TI}}_{1111}=\overline{\left(\dfrac{c_{1111}-2c_{2323}}{c_{1111}}\right)}^{\,2}
\,\,\,\overline{\left(\dfrac{1}{c_{1111}}\right)}^{\,-1}
+\overline{\left(\dfrac{4(c_{1111}-c_{2323})c_{2323}}{c_{1111}}\right)}\,,
\end{equation}
\begin{equation}
\label{eq:elast-const2}
c^{\overline{\rm
TI}}_{1122}=\overline{\left(\dfrac{c_{1111}-2c_{2323}}{c_{1111}}\right)}^{\,2}
\,\overline{\left(\dfrac{1}{c_{1111}}\right)}^{\,-1}
+\overline{\left(\dfrac{2(c_{1111}-2c_{2323})c_{2323}}{c_{1111}}\right)}\,,
\end{equation}
\begin{equation}
\label{eq:elast-const3}
c^{\overline{\rm TI}}_{1133}=\overline{\left(\dfrac{c_{1111}-2c_{2323}}{c_{1111}}\right)}\,\,
\,\,\overline{\left(\dfrac{1}{c_{1111}}\right)}^{\,-1}
\,,
\end{equation}

\begin{equation}
\label{eq:elast-const4}
c^{\overline{\rm TI}}_{2323}=\overline{\left(\dfrac{1}{c_{2323}}\right)}^{\,-1}
\,,
\end{equation}
\begin{equation}
\label{eq:elast-const5}
c^{\overline{\rm TI}}_{3333}=\overline{\left(\dfrac{1}{c_{1111}}\right)}^{\,-1}
\,,
\end{equation}
where the bar indicates an averaged value. The present expressions \eqref{eq:elast-const1}--\eqref{eq:elast-const5} are expressions (13) in the original paper~\cite{backus}, expressed in terms of Lam\'e's parameters $\lambda$ and $\mu$. The usage of the following connections: $c_{1111}=\lambda+2\mu$, $c_{2323}=\mu$ shows the equivalence of these expressions.

For small deformations we may use Backus average, but for large deformations linear elasticity is not suitable in general and we need to use a nonlinear elasticity which accounts for large deformations.
In this paper we propose an extended use of Backus average for finitely deformed materials.
A nonlinear elastic material may be treated essentially as linear elastic under small loads and small deformations.
Thus, each nonlinear elastic material may be associated with a corresponding linear elastic material.

As it can be seen from the examples considered in this paper, in general, we can observe that we cannot describe a nonlinear material fully by the information obtained from its corresponding linear elastic counterpart.
Normally, we do not have enough information for this purpose (and this should not be considered as something surprising, because nonlinear elasticity is more general and linear elasticity is a particular case of nonlinear elasticity).
In order to fill this gap in the information for modelling of the nonlinear behaviour of a material, based on information from linear elasticity, we need to use some additional information or to make an educated guess.

\section{Theoretical background}
We label a material point in the reference configuration $\mathcal{B}_\mathrm{r}$ by a position vector $\mathbf{X}$, and this point in the current configuration $\mathcal{B}$ by a position vector $\mathbf{x}$. Deformation is described by the vector field $\boldsymbol{\chi}$, which relates the position of a particle in the reference configuration to the position of the same particle in the current configuration: $\mathbf{x}=\boldsymbol{\chi}(\mathbf{X})$. The deformation gradient tensor, denoted $\mathbf{F}$, is defined by
\begin{equation}
\mathbf{F}=\Grad\boldsymbol{\chi},
\end{equation}
where $\Grad$ is the operator defined with respect to $\mathbf{X}$.

The stored energy in a transversely isotropic nonlinear elastic material can be formulated in terms of invariants $I_1$, $I_2$, $I_3$, $I_4$, $I_5$, ~\cite{Spencer}.
Thus, the strain energy per unit reference volume can be written as a function
\begin{equation*}
  W=W(I_1, I_2, I_3, I_4, I_5),
\end{equation*}
where invariants are defined by
\begin{equation*}
  I_1=\tr {\bf C}, \quad I_2=I_3\tr ({\bf C}^{-1})\,, \quad I_3=\det {\bf C}\,,
\end{equation*}

\begin{equation*}
  I_4={\bf A}\cdot{\bf CA}\,, \quad I_5={\bf A}\cdot{\bf C}^2{\bf  A}\,,
\end{equation*}
where $\bf C={\bf F}^\mathrm{T}{\bf F}$ is the right Cauchy-Green deformation tensor.

For nonlinear elastic transversely isotropic material a unit vector $\bf A$ represents a fiber direction in the reference configuration.
In our case we do not have any fibres, because, according to Backus, transverse isotropy is caused by the layered structure of the body.
Therefore, we may identify $\bf A$ with imaginary fibres or $\bf A$ can be thought of as a direction of anisotropy in the reference configuration.
In the current configuration this direction deforms to $\bf a=\bf FA$.

First Piola-Kirchhoff stress tensor can be found as a derivative of a strain energy function with respect to deformation gradient tensor
\begin{equation*}
  {\bf P}=\frac{\partial W}{\partial \bf F}\,.
\end{equation*}
Using a chain rule and the following relations
\begin{align*}
  \frac{\partial I_1}{\partial \bf F}=2 {\bf F}, \quad &\frac{\partial I_2}{\partial \bf F}=2 I_1{\bf F}-2{\bf F}{\bf F}^\mathrm{T}{\bf F}\,, \quad \frac{\partial I_3}{\partial \bf F}=2I_3{\bf F}^\mathrm{-T}\,, \\[1ex]
  &\frac{\partial I_4}{\partial {\bf F}}=2{\bf FA}\otimes{\bf A}, \quad \frac{\partial I_5}{\partial {\bf F}}=2 \left(\bf FCA\otimes A + FA \otimes CA\right)\,,
\end{align*}
we obtain
\begin{align*}
  {\bf P}=&2W_1{\bf F} + 2W_2\left(I_1 {\bf F}-{\bf F}{\bf F}^\mathrm{T} {\bf F}\right)+2W_3I_3{\bf F}^\mathrm{-T}\\
   &+2W_4{\bf FA}\otimes {\bf A}+2W_5\left(\bf FCA\otimes A + FA \otimes CA\right)\,.
\end{align*}
Thus, Cauchy stress $\boldsymbol\sigma$ can be found using a standard relation
\begin{align}\label{cauchy-stress}
  J\boldsymbol{\sigma}={\bf P}{\bf F}^\mathrm{T}=&2 W_1 {\bf B}+2W_2(I_1{\bf I}-{\bf B}){\bf B}+2W_3I_3{\bf I}\notag\\[1ex]
&+2W_4{\bf a}\otimes {\bf a}+2W_5({\bf Ba}\otimes{\bf a}+{\bf a}\otimes{\bf Ba}),
\end{align}
where $\bf B={\bf F}{\bf F}^\mathrm{T}$ is the left Cauchy-Green deformation tensor, and we remind that $J=\det \bf F$. Subscripts for energy function $W$ denote a corresponding derivative with respect to invariants $I_1$,~..., $I_5$.

In the reference configuration, where $I_1=3$, $I_2=3$, $I_3=1$, $I_4=1$, $I_5=1$, strain energy function has a zero value
\begin{equation}\label{eq:energy-zero}
  W(3,3,1,1,1)=0.
\end{equation}
Stress should also vanish in the reference configuration.
Thus, evaluating expression for stress~\eqref{cauchy-stress} at $\bf F=I$ we obtain
\begin{equation}\label{eq:stress-isotr-zero}
  W_1(3,3,1,1,1)+2W_2(3,3,1,1,1)+W_3(3,3,1,1,1)=0\,,
\end{equation}
\begin{equation}\label{eq:stress-transverse-zero}
  W_4(3,3,1,1,1)+2W_5(3,3,1,1,1)=0\,.
\end{equation}

The following conditions should be satisfied for consistency with linear transversely isotropic elasticity (Merodio \& Ogden~\cite{MO2003}):
\begin{equation}\label{eq:connection-start}
  W_{11}+4W_{12}+4W_{22}+2W_{13}+4W_{23}+W_{33}=c_{11}/4\,,
\end{equation}
\begin{equation}\label{eq:connection-c12c11}
  W_2+W_3=(c_{12}-c_{11})/4\,,
\end{equation}
\begin{equation}\label{eq:connection-c44}
  W_1+W_2+W_5=c_{44}/2\,,
\end{equation}
\begin{equation}\label{eq:connection-end1}
  W_{14}+2W_{24}+2W_{15}+W_{34}+4W_{25}+2W_{35}=(c_{13}-c_{12})/4\,,
\end{equation}
\begin{equation}\label{eq:connection-end}
  W_{44}+4W_{45}+4W_{55}+2W_5=(c_{33}-c_{11}+2c_{12}-2c_{13})/4\,.
\end{equation}
All derivatives in the expressions~\eqref{eq:connection-start}--\eqref{eq:connection-end} are evaluated in the reference configuration. Note that for these connections it is important that the vector $\mathbf{A}$ is aligned along the $X_3$ axis in the reference configuration, which is shown explicitly in Appendix \ref{app:connections}.

It is well-known that due to symmetries of Cauchy stress $\boldsymbol{\sigma}$ and small strain tensor $\boldsymbol{\varepsilon}$ the components of the elasticity tensor in linear elasticity may be written as a $6\times6$ matrix and they are present in this form in the right hand side of relations~\eqref{eq:connection-start}--\eqref{eq:connection-end}.
We can identify these components $c_{mn}$ with the components $c_{ijkl}$ of the forth-order elasticity tensor, using for following rule for indices: $(11)\rightarrow1$, $(22)\rightarrow2$, $(33)\rightarrow1$, $(23)\rightarrow4$, $(13)\rightarrow5$, $(12)\rightarrow6$, where a pair of indices in parentheses correspond to a pair of indices $ij$ and $kl$ in $c_{ijkl}$, a single index after the arrow sign corresponds to an index $m$ or $n$ in $c_{mn}$.
Thus, we can write $c_{11}=c_{1111}$, $c_{12}=c_{1122}$, $c_{33}=c_{3333}$, $c_{13}=c_{1133}$, $c_{44}=c_{2323}$.
Axis~$x_3$ represents the axis of symmetry for linear elastic transversely isotropic material.

\section{Association of linear elastic materials with specific nonlinear models}
\label{lin with nonlin models}
Following~Merodio \& Ogden \cite{MO2003} let us start with a strain energy function which depends on three invariants $W(I_1, I_3, I_4)$
\begin{equation}\label{eq:potential-134}
  W(I_1, I_3, I_4)=\hat{\mu}(I_1-3)+H(I_3)+F(I_4)\,,
\end{equation}
where $\hat{\mu}$ is a positive material constant and $F$ is a function satisfying
\newline $F(1)=0$, $F'(1)=0$.
From conditions~\eqref{eq:energy-zero} and~\eqref{eq:stress-isotr-zero} we obtain
\begin{equation*}
  H(1)=0, \quad H'(1)=-\hat{\mu}.
\end{equation*}
Condition \eqref{eq:stress-transverse-zero} is also satisfied for function \eqref{eq:potential-134}.

We may specify function $F(I_4)$ as $F(I_4)=\frac{1}{2}\alpha(I_4-1)^2$, where $\alpha$ is a positive material parameter accounting for the degree of anisotropy.
Function $F(I_4)$ is often referred to as a reinforcing model and accounts for transversely isotropic properties of a material.
Therefore, we confirm that
\begin{equation*}
F(1)=0\,, \quad   F'(1)=0,
\end{equation*}
and additionally we obtain
\begin{equation*}
F''(1)=\alpha\,.
\end{equation*}
We use the value of the second derivative $H''(1)=k$.
Thus, from expressions~\eqref{eq:connection-start}--\eqref{eq:connection-end} we obtain

\begin{align}\label{eq:system}
  &c_{11}=4k, \quad c_{33}=4(k+\alpha), \quad c_{44}=2\hat{\mu},\quad c_{13}(=c_{12})=4(k-\hat{\mu}).
\end{align}

Note that from the strain energy potential~\eqref{eq:potential-134} it is always possible to find elastic constants $c_{11}$\,,...,\,$c_{44}$.
In contrast, in the present paper we are interested in the opposite problem. From Backus averaging we want to find parameters $\hat{\mu}$, $\alpha$ and $k$ so that they would help to define function~\eqref{eq:potential-134}.
Thus, from function \eqref{eq:potential-134} we will be able to predict overall transversely isotropic mechanical behaviour of layers as a whole structure under significant loads and thus experiencing large deformations.
Note that in this case relations~\eqref{eq:system} can be viewed as a system of $5$ linear algebraic equations with $3$ unknown variables $\hat{\mu}$, $\alpha$ and $k$, and thus it may have no solutions for some combinations of elastic constants $c_{11}$\,,...,\,$c_{44}$.
The system~\eqref{eq:system} will always have a solution if the additional condition for elastic constants is satisfied
\begin{equation*}
  c_{12}=c_{13}=c_{11}-2c_{44}.
\end{equation*}

Let us consider another strain energy function with the reinforcing model defined by $F(I_5)=\frac{1}{2}\gamma(I_5-1)^2$
\begin{equation}\label{eq:potential-135}
W(I_1, I_3, I_5)=\hat{\mu}(I_1-3)+H(I_3)+F(I_5)\,.
\end{equation}
 Thus, we have
\begin{equation*}
F(1)=0\,, \quad   F'(1)=0\,, \quad F''(1)=\gamma\,.
\end{equation*}
Using connections~\eqref{eq:connection-start}--\eqref{eq:connection-end} we obtain
\begin{align}\label{eq:system-I-5}
  &c_{11}=4k, \quad c_{33}=4(k+\gamma), \quad c_{44}=2\hat{\mu},\quad c_{13}(=c_{12})=4(k-\hat{\mu}).
\end{align}
Note that if we take material parameters $\alpha=\gamma$ in expressions~\eqref{eq:potential-134} and~\eqref{eq:potential-135}, then using connections ~\eqref{eq:connection-start}--\eqref{eq:connection-end} we obtain the same linear elastic material corresponding to different nonlinear elastic materials (due to different invariants $I_4$ and $I_5$ in the reinforcing models).
Thus, two different nonlinear elastic materials may correspond to the same linear elastic material. It is important to bear in mind this fact: in order to properly model a nonlinear elastic material we need to use additional information about a possible behaviour of a such material in nonlinear regime.
Invariants $I_4$ and $I_5$ featuring in expressions~\eqref{eq:potential-134} and \eqref{eq:potential-135} model differently the behaviour of a material under large deformations.
Invariant $I_5$ accounts for an additional reinforcement under shear deformations. For details we refer to Merodio \& Ogden~\cite{MO2005}.

Let us consider another example. We define strain energy function as
\begin{equation*}
  W=\hat{\mu}(I_1-3)+K(I_2)+H(I_3)+F(I_4).
\end{equation*}
Functions $K(I_2)$, $H(I_3)$ and $F(I_4)$ are defined so that the following conditions are satisfied
\begin{equation*}
  K(3)=0\,, \quad K''(3)=q\,, \quad H(1)=0\,, \quad H'(1)=d\,, \quad H''(1)=k\,,
  \end{equation*}
\begin{equation*}
\quad F(1)=0\,, \quad F'(1)=0\,, \quad F''(1)=\alpha\,.
\end{equation*}
From condition~\eqref{eq:stress-isotr-zero} we obtain
\begin{equation*}
  K'(3)=-(\hat{\mu}+d)/2.
\end{equation*}
From connections~\eqref{eq:connection-start}--\eqref{eq:connection-end} we obtain
\begin{equation*}
  4q+k=c_{11}/4\,, \quad d-\hat{\mu}=(c_{12}-c_{11})/2\,, \quad \hat{\mu}+d=c_{44}, \quad \alpha=(c_{33}-c_{11})/4\,.
\end{equation*}
Note that parameters $q$ and $k$ are not uniquely defined in this case. Thus, again in order to extend modelling for nonlinear regime we need to make some educated guess or use some additional information about material working under large loads.

At the end of this section lets us consider the potential with a specified function $H(I_3)$ depending on the volumetric deformation captured by the invariant $I_3$.
For example, we may consider the following potential
\begin{equation}\label{eq:poten-I4}
  W=\frac{\mu}{2}(I_1-3)-\mu\log \sqrt{I_3}+\frac{\lambda}{2}(\log\sqrt{I_3})^2+\frac{1}{2}\alpha(I_4-1)^2\,,
\end{equation}
where $\lambda$ and $\mu$ are Lam\'e's parameters.
Note that for this potential conditions~\eqref{eq:stress-isotr-zero} and~\eqref{eq:stress-transverse-zero} are satisfied. We want to find material parameters $\mu$, $\lambda$ and $\alpha$ in \eqref{eq:poten-I4} from the connections~\eqref{eq:connection-start}--\eqref{eq:connection-end}.  Thus, we have
\begin{equation*}
  2\mu+\lambda=c_{11}, \quad \mu=c_{44}, \quad \mu=(c_{11}-c_{12})/4, \quad c_{13}=c_{12}, \quad \alpha=(c_{33}-c_{11})/4.
\end{equation*}
We note that for compatibility we need an additional condition to be satisfied $c_{44}=(c_{11}-c_{12})/2$.

Let us consider a particular case of expression \eqref{eq:poten-I4} without the term containing invariant $I_4$. Thus, we have the following expression for the potential
\begin{equation}\label{eq:iso-poten}
  W=\frac{\mu}{2}(I_1-3)-\mu\log \sqrt{I_3}+\frac{\lambda}{2}(\log\sqrt{I_3})^2\,,
\end{equation}
typically used for modelling isotropic nonlinear elastic materials~(Bonet \& Wood~\cite{Bonet2008}). It is worth noting that from the expressions ~\eqref{eq:connection-start}--\eqref{eq:connection-end1} we correctly recover expressions of the components of linear elastic tensor $c_{ijkl}=\lambda\delta_{ij}\delta_{kl}+\mu(\delta_{ik}\delta_{jl}+\delta_{il}\delta_{jk})$ in terms of Lam\'e's parameters:
\begin{equation*}
  c_{1111}=2\mu+\lambda, \quad c_{2323}=\mu, \quad c_{1122}=\lambda, \quad c_{1133}=\lambda.
\end{equation*}

For incompressible linear elastic material the number of elastic constants reduces from five to three (Merodio \& Ogden~\cite{MO2005}). For this case connections~\eqref{eq:connection-start}--\eqref{eq:connection-end} can be specialized appropriately. For details we refer to Merodio \& Ogden~\cite{MO2005}.
\section{Formulation based on a predeformed configuration}
\label{predeformed-config}
In this section we use a different approach and explore the possibility of using a prestress formulation for Backus averaging for finitely deformed materials of a layered structure with waves being treated as superimposed displacements on a current deformed configuration.

%%%%%%%%%%%%%%%%%%%%%%%%%%%%%%%%%%%
Backus (1962) showed that a stack of thin isotropic layers can be considered equivalently as a homogenized transversely isotropic media. All derivations were done within the context of a linear elastic theory. In this section we attempt to see if a similar averaging can be made within finite elasticity for large deformations using a prestress formulation.

In Backus averaging the load applied to the horizontal plane of a medium which consists of thin horizontal layers can be reasonably considered as prestress.  The wave propagating in this media can be considered as a displacement field, superimposed on the underlying loaded current configuration. Although Backus (1962) considers a wave propagation, his problem is essentially static. In his paper displacement fields do not depend on time. We also consider incremental displacement field which is static and does not depend on time.

In this section instead of the first Piola-Kirchhoff stress tensor we use a nominal stress tensor, which is a transpose of the first Piola-Kirchhoff stress tensor $\mathbf{T}=\mathbf{P}^{\mathrm{T}}$.

Let us consider a small displacement superimposed on a finitely deformed configuration
\begin{equation*}
  \dot{\bf x}=\dot{\boldsymbol{\chi}}({\bf X})\,.
\end{equation*}
We use notation $\mathbf{u}=\dot{\bf x}$ for the incremental displacement $\dot{\bf x}$. We define that the increment in the deformation gradient is
\begin{equation*}
\dot{\mathbf{F}}=\Grad \dot{\boldsymbol \chi}.
\end{equation*}
Treating $\bf u$ as a function of $\bf x$, we obtain
\begin{equation}\label{dot-F}
  \dot{\mathbf{F}}=(\grad \bf u) \,\bf F.
\end{equation}
Note that operator $\grad$ is defined with respect to $\bf x$, and we denote $\mathbf{L}=\grad\mathbf{u}$.
Incrementing expression
\begin{equation}\label{Cauchy-Nominal}
  J\boldsymbol{\sigma}=\mathbf{F}\mathbf{T}
\end{equation}
for compressible case, we obtain
\begin{equation}
\dot{J}\boldsymbol{\sigma}+J\dot{\boldsymbol{\sigma}}=\dot{\mathbf{F}}\mathbf{T}+\mathbf{F}\dot{\mathbf{T}}.
\end{equation}
Rearranging,
\begin{equation}
\dot{\boldsymbol{\sigma}}=J^{-1}\dot{\mathbf{F}}\mathbf{T}+J^{-1}\mathbf{F}\dot{\mathbf{T}}-J^{-1}\dot{J}\boldsymbol{\sigma}.
\end{equation}
We use expression \eqref{dot-F}, expression $\dot{J}=J\tr\mathbf{L}$ and a push forward version of increment in nominal stress tensor
\begin{equation}\label{push-f-inc-nom}
\dot{\mathbf{T}}_0=J^{-1}\mathbf{F}\dot{\mathbf{T}}.
\end{equation}
For details we refer to the book by R.W. Ogden ~\cite{Ogden}, Chapter 6. Thus, we obtain
\begin{equation}\label{compres}
\dot{\boldsymbol{\sigma}}=\mathbf{L}\boldsymbol{\sigma}+\dot{\mathbf{T}}_0-(\tr\mathbf{L})\boldsymbol{\sigma},
\end{equation}
or we can rewrite
\begin{equation}\label{re-compres}
\dot{\boldsymbol{\sigma}}=[\mathbf{L}-(\tr\mathbf{L})\mathbf{I}]\boldsymbol{\sigma}+\dot{\mathbf{T}}_0.
\end{equation}
For incompressible case $\tr\mathbf{L}=0$,
\begin{equation}\label{incomp}
\dot{\boldsymbol{\sigma}}=\mathbf{L}\boldsymbol{\sigma}+\dot{\mathbf{T}}_0.
\end{equation}
Increment in Cauchy stress $\dot{\boldsymbol{\sigma}}$ is a symmetric tensor.

Note that $\mathbf{L}$ and $\dot{\mathbf{T}}_0$ are not symmetric, but $\dot{\boldsymbol{\sigma}}$ is a symmetric tensor. The proof follows from the expression \eqref{updated-rot-bal}, given in Appendix \ref{app:incremental}, and also from expression~(6.2.21) in~\cite{Ogden}, given in a different notation. 
Expression~(6.2.21) in the present notation can be written as
\begin{equation}
  \mathbf{L}\boldsymbol{\sigma}+\dot{\mathbf{T}}_0=\dot{\mathbf{T}}^{\mathrm{T}}_0+\boldsymbol{\sigma}\mathbf{L}^{\mathrm{T}}.
\end{equation}
Thus, symmetry of \eqref{incomp} is established. For compressible case~\eqref{compres}, increment $\dot{\boldsymbol{\sigma}}$ is also symmetric, since
$(\tr\mathbf{L})\boldsymbol{\sigma}$ is symmetric.

Using incremental constitutive law for compressible case
\begin{equation}\label{incr-constit-law}
\mathbf{\dot{T}}_0=\boldsymbol{\mathcal{A}}_0\mathbf{L}.
\end{equation}
Substituting \eqref{incr-constit-law} in \eqref{re-compres}, we obtain
\begin{equation}\label{fin-incr-law}
\dot{\boldsymbol{\sigma}}=[\mathbf{L}-(\tr\mathbf{L})\mathbf{I}]\boldsymbol{\sigma}+\boldsymbol{\mathcal{A}}_0\mathbf{L}.
\end{equation}
For isotropic material non-zero instantaneous elastic moduli $\boldsymbol{\mathcal{A}}_0$ can be specified, see Chapter 6 in~\cite{Ogden}. Thus, we can obtain in components from \eqref{fin-incr-law}
\begin{align}
\label{index-begin-incr-law}
  \dot{\sigma}_{11}&=\frac{\partial u_1}{\partial x_1}\sigma_{11}+\frac{\partial u_1}{\partial x_2}\sigma_{21}+\frac{\partial u_1}{\partial x_3}\sigma_{31}-\left(\frac{\partial u_1}{\partial x_1}+\frac{\partial u_2}{\partial x_2}+\frac{\partial u_3}{\partial x_3}\right)\sigma_{11}\nonumber\\
  &+\mathcal{A}_{01111}\frac{\partial u_1}{\partial x_1}+\mathcal{A}_{01122}\frac{\partial u_2}{\partial x_2}+\mathcal{A}_{01133}\frac{\partial u_3}{\partial x_3},
  \end{align}
\begin{align}
  \dot{\sigma}_{22}&=\frac{\partial u_2}{\partial x_1}\sigma_{12}+\frac{\partial u_2}{\partial x_2}\sigma_{22}+\frac{\partial u_2}{\partial x_3}\sigma_{32}-\left(\frac{\partial u_1}{\partial x_1}+\frac{\partial u_2}{\partial x_2}+\frac{\partial u_3}{\partial x_3}\right)\sigma_{22}\nonumber\\
  &+\mathcal{A}_{02211}\frac{\partial u_1}{\partial x_1}+\mathcal{A}_{02222}\frac{\partial u_2}{\partial x_2}+\mathcal{A}_{02233}\frac{\partial u_3}{\partial x_3},
  \end{align}
  \begin{align}
  \dot{\sigma}_{33}&=\frac{\partial u_3}{\partial x_1}\sigma_{13}+\frac{\partial u_3}{\partial x_2}\sigma_{23}+\frac{\partial u_3}{\partial x_3}\sigma_{33}-\left(\frac{\partial u_1}{\partial x_1}+\frac{\partial u_2}{\partial x_2}+\frac{\partial u_3}{\partial x_3}\right)\sigma_{33}\nonumber\\
  &+\mathcal{A}_{03311}\frac{\partial u_1}{\partial x_1}+\mathcal{A}_{03322}\frac{\partial u_2}{\partial x_2}+\mathcal{A}_{03333}\frac{\partial u_3}{\partial x_3},
  \end{align}
  \begin{align}
  \dot{\sigma}_{13}&=\frac{\partial u_1}{\partial x_1}\sigma_{13}+\frac{\partial u_1}{\partial x_2}\sigma_{23}+\frac{\partial u_1}{\partial x_3}\sigma_{33}-\left(\frac{\partial u_1}{\partial x_1}+\frac{\partial u_2}{\partial x_2}+\frac{\partial u_3}{\partial x_3}\right)\sigma_{13}\nonumber\\
  &+\mathcal{A}_{01313}\frac{\partial u_3}{\partial x_1}+\mathcal{A}_{01331}\frac{\partial u_1}{\partial x_3},
  \end{align}
   \begin{align}
  \dot{\sigma}_{23}&=\frac{\partial u_2}{\partial x_1}\sigma_{13}+\frac{\partial u_2}{\partial x_2}\sigma_{23}+\frac{\partial u_2}{\partial x_3}\sigma_{33}-\left(\frac{\partial u_1}{\partial x_1}+\frac{\partial u_2}{\partial x_2}+\frac{\partial u_3}{\partial x_3}\right)\sigma_{23}\nonumber\\
  &+\mathcal{A}_{02323}\frac{\partial u_3}{\partial x_2}+\mathcal{A}_{02332}\frac{\partial u_2}{\partial x_3},
  \end{align}
  \begin{align}
  \label{index-last-incr-law}
  \dot{\sigma}_{12}&=\frac{\partial u_1}{\partial x_1}\sigma_{12}+\frac{\partial u_1}{\partial x_2}\sigma_{22}+\frac{\partial u_1}{\partial x_3}\sigma_{32}-\left(\frac{\partial u_1}{\partial x_1}+\frac{\partial u_2}{\partial x_2}+\frac{\partial u_3}{\partial x_3}\right)\sigma_{12}\nonumber\\
  &+\mathcal{A}_{01212}\frac{\partial u_2}{\partial x_1}+\mathcal{A}_{01221}\frac{\partial u_1}{\partial x_2}.
  \end{align}
Now we want to follow similar procedures which were done for linear elastic case in~\cite{backus} and in~\cite{Slawinski}, Section $4.2$, i.e.
we need to rearrange terms in expressions~\eqref{index-begin-incr-law}--\eqref{index-last-incr-law} according to how they vary through out the hight of the stack of layers: slowly or fast. Terms which vary significantly along the hight of the stack of layers should be brought to one side. According to Backus procedure on the other side we may have slow-varying terms, multiplied by the abruptly varying terms. Due to complexity of expressions~\eqref{index-begin-incr-law}--\eqref{index-last-incr-law}, apparently, it is impossible to follow the similar procedures done by Backus for linear elastic case. Also, note that in the present setting due to the superimposed displacements on the current deformed configuration we have incremental Cauchy stress $\dot{\boldsymbol{\sigma}}$, which is absent in Backus formulation due to his solution obtained purely within linear elastic theory.
Also, it is worth noting that here instantaneous elastic moduli $\boldsymbol{\mathcal{A}}_0$ are functions of strain, whereas in a classical linear elastic case elastic parameters are simply constants.
\section{Summary and Conclusions}
In this paper we proposed a new approach for calculating the overall properties of a homogenized material. This approach is based on the combination of linear and non-linear elastic theories. Let us summarise this procedure. For each isotropic layer we know elasticity constants $c_{1111}$ and $c_{2323}$, or equivalently Lam\'e's parameters $\lambda$ and $\mu$. Then, we use expressions \eqref{eq:elast-const1}--\eqref{eq:elast-const5}, and we find elastic constants for overall material (but still in linear regime for small deformations). Then, we use connections \eqref{eq:connection-start}--\eqref{eq:connection-end} and construct the stain energy potential, which accounts for the behaviour of the material in the nonlinear regime. In Section \ref{lin with nonlin models} we give examples of such potentials, including a fully specified potential (20). We note that although, in general, we need some additional information on how material works in non-linear regime, a potential with the standard reinforcing model may be used by default.

We think it is also instructive to show another approach described in Section \ref{predeformed-config}, despite the fact that it did not lead to a successful result. Nonetheless, we think that sometimes a negative result should be shown too. If one compares equations given in book ~\cite{Slawinski}, Section 4.2 and those presented in this paper \eqref{index-begin-incr-law}--\eqref{index-last-incr-law}, we will find many similar terms, elastic moduli $\boldsymbol{\mathcal{A}}_0$ now depend on the strain, they are not constants as it is the case for linear elastic problem. As mentioned before, due to complexity of expressions \eqref{index-begin-incr-law}--\eqref{index-last-incr-law}, apparently, a further progression becomes impossible. The derived equations can be also used for modelling small deformations superimposed on a deformed configuration. If a deformation field depends on time, we may think about the deformation field as a wave propagating in a prestressed media. In Section 4 we concluded that tensor $\dot{\boldsymbol{\sigma}}$ is symmetric, and thus we need to consider only 6 expressions \eqref{index-begin-incr-law}--\eqref{index-last-incr-law}. Symmetry of tensor $\dot{\boldsymbol{\sigma}}$ follows from rotational balance equation by taking increments. In turn,  rotational balance equation is a consequence of objectivity of a constitutive law, and the proof of this fact is given in Appendix \ref{app:incremental}.

%%%%%%%%%%%%%%%%%%%%%%%%%%%%
\section*{Acknowledgments}
%%%%%%%%%%%%%%%%%%%%%%%%%%%%
I would like to thank and acknowledge discussions with Professor Ray W. Ogden and Professor Michael Slawinski.
This research was performed in the context of The Geomechanics Project supported by Husky Energy.
Also, this research was partially supported by the Natural Sciences and Engineering Research Council of Canada, grant 238416-2013.

%%%%%%%%%%%%%%%%%%%%%%%%%%%%
\appendix
\numberwithin{equation}{section}

%%%%%%%%%%%%%%%%%%%%%%%%%%%%%%%%%%%%%%%%%%

\section{Connections between nonlinear theory and the classical linear elastic transversely isotropic elasticity}
\label{app:connections}
%%%%%%%%%%%%%%%%%%%%%%%%%%%%%%%%%%%%%%%%%%
%In what follows the convention for switching indices after differentiation of a scalar function was not used.
The first derivative of the strain energy function $W(I_1, I_2, I_3, I_4, I_5)$ with respect to $\mathbf{F}$ is
\begin{eqnarray}
\frac{\partial W}{\partial F_{i \alpha}}&=&2 W_1 F_{i \alpha}+2W_2(I_1F_{i\alpha}-F_{i\gamma}F_{k\gamma}F_{k\alpha})+2W_3I_3F^{-1}_{\alpha i} \notag\\
&+&2W_4F_{i\tau}A_\tau A_{\alpha}+2W_5(F_{i\pi}F_{k\pi}F_{k\rho} A_\rho A_{\alpha}+F_{i\gamma}A_{\gamma}F_{p\alpha}F_{p\omega}A_\omega).
\end{eqnarray}
We have used
\begin{equation*}\label{}
  \frac{\partial I_1}{\partial F_{i \alpha}}=2 F_{i \alpha}, \quad \frac{\partial I_2}{\partial F_{i \alpha}}=2 (I_1F_{i \alpha}-F_{i\gamma}F_{k\gamma}F_{k\alpha}), \quad \frac{\partial I_3}{\partial F_{i \alpha}}=2I_3F^{-1}_{\alpha i},
\end{equation*}
\begin{equation*}
  \frac{\partial I_4}{\partial F_{i \alpha}}=2F_{i\tau} A_\tau A_{\alpha}, \quad \frac{\partial I_5}{\partial F_{i \alpha}}=2(F_{i\pi}F_{k\pi}F_{k\rho}A_\rho A_{\alpha}+F_{i\gamma}A_{\gamma}F_{p\alpha}F_{p\omega}A_\omega).
\end{equation*}
Components of the elastic moduli tensor associated with the conjugate pair $(\mathbf{P}, \mathbf{F})$, which are a second derivative with respect to $\mathbf{F}$, can be obtained as
\begin{eqnarray}\label{second-der}
\mathcal{A}_{j \beta i \alpha}&=&
2 \Big[W_{11} 2 F_{j \beta}+2W_{12}(I_1F_{j\beta}-F_{j\gamma}F_{k\gamma}F_{k\beta})+2W_{13}I_3F^{-1}_{\beta j}+W_{14}2F_{j\tau}A_\tau A_{\beta}\notag\\
&+&2W_{15}(F_{j\pi}F_{k\pi}F_{k\rho}A_\rho A_{\beta}+F_{j\gamma}A_{\gamma}F_{p\beta}F_{p\omega} A_\omega)\Big]F_{i\alpha}+2W_1\frac{\partial F_{i\alpha}}{\partial F_{j\beta}}\notag\\
&+&2 \Big[W_{21} 2 F_{j \beta}+2W_{22}(I_1F_{j\beta}-F_{j\gamma}F_{k\gamma}F_{k\beta})+2W_{23}I_3F^{-1}_{\beta j}+W_{24}2F_{j\tau}A_\tau A_{\beta}\notag\\
&+&2W_{25}(F_{j\pi}F_{k\pi}F_{k\rho}A_\rho A_{\beta}+F_{j\gamma}A_{\gamma}F_{p\beta}F_{p\omega} A_\omega)\Big](I_1F_{i\alpha}-F_{i\gamma}F_{k\gamma}F_{k\alpha})\notag\\
&+&2W_2\Big(\frac{\partial I_1}{\partial F_{j\beta}}F_{i\alpha}+I_1\frac{\partial F_{i\alpha}}{\partial F_{j \beta}}-\frac{\partial F_{i\gamma}}{\partial F_{j\beta}}F_{k\gamma}F_{k\alpha}-F_{i\gamma}\frac{\partial F_{k\gamma}}{\partial F_{j\beta}}F_{k\alpha}-F_{i\gamma}F_{k\gamma}\frac{\partial F_{k\alpha}}{\partial F_{j \beta}}\Big)\notag\\
&+&2\Big[W_{31} 2 F_{j \beta}+2W_{32}(I_1F_{j\beta}-F_{j\gamma}F_{k\gamma}F_{k\beta})+2W_{33}I_3F^{-1}_{\beta j}+W_{34}2F_{j\tau}A_\tau A_{\beta}\notag\\
&+&2W_{35}(F_{j\pi}F_{k\pi}F_{k\rho}A_\rho A_{\beta}+F_{j\gamma}A_{\gamma}F_{p\beta}F_{p\omega} A_\omega)\Big]I_3 F^{-1}_{\alpha i}+2W_3\Big(\frac{\partial I_3}{\partial F_{j \beta}}F^{-1}_{\alpha i}+I_3\frac{\partial F^{-1}_{\alpha i}}{\partial F_{j \beta}}\Big)\notag\\
&+&2 \Big[W_{41} 2 F_{j \beta}+2W_{42}(I_1F_{j\beta}-F_{j\gamma}F_{k\gamma}F_{k\beta})+2W_{43}I_3F^{-1}_{\beta j}+W_{44}2F_{j\tau}A_\tau A_{\beta}\notag\\
&+&2W_{45}(F_{j\pi}F_{k\pi}F_{k\rho}A_\rho A_{\beta}+F_{j\gamma}A_{\gamma}F_{p\beta}F_{p\omega} A_\omega)\Big]F_{i\tau}A_\tau A_{\alpha}+2W_4\frac{\partial F_{i\delta}}{\partial F_{j\beta}}A_\delta A_{\alpha}\notag\\
&+&2\Big[W_{51} 2 F_{j \beta}+2W_{52}(I_1F_{j\beta}-F_{j\gamma}F_{k\gamma}F_{k\beta})+2W_{53}I_3F^{-1}_{\beta j}+W_{54}2F_{j\tau}A_\tau A_{\beta}\notag\\
&+&2W_{55}(F_{j\pi}F_{k\pi}F_{k\rho}A_\rho A_{\beta}+F_{j\gamma}A_{\gamma}F_{p\beta}F_{p\omega} A_\omega)\Big](F_{i\pi}F_{k\pi}F_{k\rho}A_\rho A_{\alpha}+F_{i\gamma}A_{\gamma}F_{p\alpha}F_{p\omega}A_\omega)\notag\\
&+&2W_5\Big(\frac{\partial F_{i\pi}}{\partial F_{j\beta}}F_{k\pi}F_{k\rho}A_\rho A_\alpha+F_{i\pi}\frac{\partial F_{k\pi}}{\partial F_{j\beta}}F_{k\rho}A_\rho A_{\alpha}+F_{i\pi}F_{k\pi}\frac{\partial F_{k\rho}}{\partial F_{j\beta}}A_\rho A_{\alpha}\notag\\
&+&\frac{\partial F_{i\gamma}}{\partial F_{j\beta}}A_{\gamma}F_{p\alpha}F_{p\omega}A_\omega+F_{i\gamma}A_{\gamma}\frac{\partial F_{p\alpha}}{\partial F_{j\beta}}F_{p\omega}A_{\omega}+F_{i\gamma}A_{\gamma}F_{p\alpha}\frac{\partial F_{p\omega}}{\partial F_{j\beta}}A_\omega\Big).
\end{eqnarray}

We use
\begin{equation*}
  \frac{\partial F^{-1}_{\alpha i}}{\partial F_{j \beta}}=-F^{-1}_{\alpha j}F^{-1}_{\beta i},
\end{equation*}
and evaluate expression \eqref{second-der} in the reference configuration, where $I_1=3$, $I_3=1$, $F_{i\alpha}=\delta_{i\alpha}$. Thus, we obtain
\begin{eqnarray}\label{evaluation-ref}
  \mathcal{A}_{j\beta i\alpha}&=&2[W_{11}2\delta_{j\beta}+4W_{12}\delta_{j\beta}+2W_{13}\delta_{j\beta }+2W_{14}A_jA_\beta+4W_{15}A_jA_\beta]\delta_{i\alpha}\notag\\
 &+&2W_1\delta_{ij}\delta_{\alpha\beta}+2[W_{21}2\delta_{j\beta}+4W_{22}\delta_{j\beta}+2W_{23}\delta_{j\beta }+2W_{24}A_jA_\beta+4W_{25}A_jA_\beta]2\delta_{i\alpha}\notag\\
 &+&2W_2[2\delta_{j\beta}\delta_{i\alpha}+\delta_{ij}\delta_{\alpha\beta}-\delta_{i\beta}\delta_{j\alpha}]\notag\\
 &+&2[W_{31}2\delta_{j\beta}+4W_{32}\delta_{j\beta}+2W_{33}\delta_{j\beta}+2W_{34}A_jA_\beta+4W_{35}A_jA_\beta]\delta_{\alpha i}\notag\\
 &+&2W_3[2\delta_{\beta j}\delta_{\alpha i}-\delta_{\alpha j}\delta_{\beta i}]\notag\\
 &+&2[W_{41}2\delta_{j\beta}+4W_{42}\delta_{j\beta}+2W_{43}\delta_{j\beta}+2W_{44}A_jA_\beta+4W_{45}A_jA_\beta]A_iA_\alpha\notag\\
 &+&2W_4\delta_{ij}A_\beta A_{\alpha}\notag\\
 &+&4[W_{51}2\delta_{j\beta}+4W_{52}\delta_{j\beta}+2W_{53}\delta_{j\beta}+2W_{54}A_jA_\beta+4W_{55}A_jA_\beta]A_iA_\alpha\notag\\
 &+&2W_5(3\delta_{ij}A_{\alpha}A_\beta+\delta_{i\beta}A_jA_\alpha+A_i A_j\delta_{\alpha\beta}+\delta_{\alpha j}A_i A_\beta).
\end{eqnarray}

Since $X_3$ is the axis of symmetry of a linear elastic transversely isotropic material, a unit vector~$\bf{A}$ should be aligned along the axis $X_3$ in the reference configuration. Thus, $A_1=A_2=0, A_3=1$.
Therefore, we have
\begin{eqnarray*}
  \mathcal{A}_{1111} &=& 2[2W_{11}+4W_{12}+2W_{13}]+2W_1 \\
   &+& 4[2W_{21}+4W_{22}+2W_{23}]+ 4W_2 \\
  &+& 2[2W_{31}+4W_{32}+2W_{33}]+2W_3 . \\
\end{eqnarray*}
Taking into consideration expression \eqref{eq:stress-isotr-zero}, we obtain
\begin{equation}\label{}
  \mathcal{A}_{1111}(=c_{11})=4W_{11}+16W_{12}+8W_{13}+16W_{22}+16W_{23}+4W_{33},
\end{equation}
which is the expression \eqref{eq:connection-start} in the paper. All remaining connections \eqref{eq:connection-c12c11}--\eqref{eq:connection-end}
can be obtained from \ref{evaluation-ref}, using \eqref{eq:stress-isotr-zero}, \eqref{eq:stress-transverse-zero}.

%%%%%%%%%%%%%%%%%%%%%%%%%%
\section{}
\label{app:incremental}
%%%%%%%%%%%%%%%%%%%%%%%%%%

We increment rotational balance equation
\begin{equation}\label{rot-balance}
  \bf F T=T^{\mathrm{T}} F^{\mathrm{T}}.
\end{equation}
Thus we have
\begin{equation*}
  \dot{\mathbf{F}}\mathbf{T}+\mathbf{F}\dot{\mathbf{T}}=\dot{\mathbf{T}}^{\mathrm{T}}\mathbf{F}^{\mathrm{T}}+\mathbf{T}^{\mathrm{T}}\dot{\mathbf{F}}^{\mathrm{T}}
\end{equation*}
Using expressions \eqref{dot-F}, \eqref{push-f-inc-nom} and \eqref{Cauchy-Nominal} we obtain
\begin{equation}\label{updated-rot-bal}
  \mathbf{L}\boldsymbol{\sigma}+\dot{\mathbf{T}}_0=\dot{\mathbf{T}}^{\mathrm{T}}_0+\boldsymbol{\sigma}\mathbf{L}^{\mathrm{T}}.
\end{equation}
It is also worth to mention here that rotational balance equation is a consequence of objectivity of the constitutive law.
Let us show that the term $\mathbf{F}\frac{\partial W}{\partial\mathbf{F}}$ is symmetric, which is a consequence of objectivity of the constitutive law. The standard requirements of objectivity show that $W$ depends on $\mathbf{F}$ only through $\mathbf{C}=\mathbf{F}^\mathrm{T}\mathbf{F}$.
Since the function $W$ is objective we have
\begin{equation}
\label{objectivity}
W(\mathbf{Q}\mathbf{F})=W(\mathbf{F}).
\end{equation}
Using polar decomposition $\mathbf{F}=\mathbf{R}\mathbf{U}$ and taking $\mathbf{R}=\mathbf{Q}^\mathrm{T}$, we can rewrite the left hand side of (\ref{objectivity}) as
\begin{equation}
W(\mathbf{U})=W(\mathbf{F}).
\end{equation}
Since $\mathbf{U}^2=\mathbf{C}$ we have
\begin{equation}
W(\mathbf{F})=W(\mathbf{U})=W(\mathbf{C}^{1/2})=\bar{W}(\mathbf{C}).
\end{equation}
Using the chain rule we have
\begin{equation}
\mathbf{F}\frac{\partial W(\mathbf{F})}{\partial\mathbf{F}}=\mathbf{F}\frac{\partial\bar{W}(\mathbf{C})}{\partial\mathbf{F}}=\mathbf{F}\frac{\partial\bar{W}}{\partial\mathbf{C}}\frac{\partial\mathbf{C}}{\partial\mathbf{F}}.
\end{equation}
We will show that the term $\mathbf{F}\frac{\partial\bar{W}}{\partial\mathbf{C}}\frac{\partial\mathbf{C}}{\partial\mathbf{F}}$ is symmetric which ensures the symmetry of $\mathbf{F}\frac{\partial W(\mathbf{F})}{\partial\mathbf{F}}$.
In components we differentiate
\begin{equation}
\label{main_expression}
F_{i\alpha}\left(\frac{\partial\bar{W}}{\partial\mathbf{F}}\right)_{\alpha j}=F_{i\alpha}\frac{\partial\bar{W}}{\partial C_{\beta\gamma}}\frac{\partial C_{\beta\gamma}}{\partial F_{j\alpha}}.
\end{equation}
We will calculate separately
\begin{equation}
\frac{\partial C_{\beta\gamma}}{\partial F_{j\alpha}}=\frac{\partial}{\partial F_{j\alpha}}(F_{p\beta}F_{p\gamma})=\delta_{jp}\delta_{\alpha\beta}F_{p\gamma}+F_{p\beta}\delta_{jp}\delta_{\alpha\gamma}=\delta_{\alpha\beta}F_{j\gamma}+\delta_{\alpha\gamma}F_{j\beta}.
\end{equation}
Therefore, relation (\ref{main_expression}) can be rewritten as
\begin{equation}
\label{calc1}
F_{i\alpha}\frac{\partial\bar{W}}{\partial C_{\beta\gamma}}(\delta_{\alpha\beta}F_{j\gamma}+\delta_{\alpha\gamma}F_{j\beta})=F_{i\beta}F_{j\gamma}\frac{\partial\bar{W}}{\partial C_{\beta\gamma}}+F_{i\gamma}F_{j\beta}\frac{\partial\bar{W}}{\partial C_{\beta\gamma}}.
\end{equation}
Since $\beta$ and $\gamma$ are indices of summation, we can switch them in the first term of (\ref{calc1}), therefore,
\begin{equation}
F_{i\beta}F_{j\gamma}\frac{\partial\bar{W}}{\partial C_{\beta\gamma}}+F_{i\gamma}F_{j\beta}\frac{\partial\bar{W}}{\partial C_{\beta\gamma}}=F_{i\gamma}F_{j\beta}\frac{\partial\bar{W}}{\partial C_{\gamma\beta}}+F_{i\gamma}F_{j\beta}\frac{\partial\bar{W}}{\partial C_{\beta\gamma}}
\end{equation}
Since tensor $\mathbf{C}$ is symmetric
\begin{align}
&F_{i\gamma}F_{j\beta}\frac{\partial\bar{W}}{\partial C_{\gamma\beta}}+F_{i\gamma}F_{j\beta}\frac{\partial\bar{W}}{\partial C_{\beta\gamma}}= \\ \nonumber
&=F_{i\gamma}F_{j\beta}\frac{\partial\bar{W}}{\partial C_{\beta\gamma}}+F_{i\gamma}F_{j\beta}\frac{\partial\bar{W}}{\partial C_{\beta\gamma}}=2F_{i\gamma}F_{j\beta}\frac{\partial\bar{W}}{\partial C_{\beta\gamma}}=2F_{i\gamma}F_{j\beta}\left(\frac{\partial\bar{W}}{\partial \mathbf{C}}\right)_{\gamma\beta}.
\end{align}
In direct notation we have finally
\begin{equation}
\mathbf{F}\frac{\partial\bar{W}}{\partial\mathbf{C}}\frac{\partial\mathbf{C}}{\partial\mathbf{F}}=2\mathbf{F}\frac{\partial\bar{W}}{\partial\mathbf{C}}\mathbf{F}^T,
\end{equation}
which proves that the expression $\mathbf{F}\frac{\partial W(\mathbf{F})}{\partial\mathbf{F}}$ is symmetric.
%Note that we used a convention that
%\begin{equation}
%\left(\frac{\partial W}{\partial \mathbf{C}}\right)_{\gamma\beta}=\frac{\partial W}{\partial C_{\beta\gamma}}.
%\end{equation}
%Here we have to note that since tensor $\mathbf{C}$ is symmetric in this particular problem this convention is not important.

\bibliographystyle{unsrt}
\bibliography{MS}
%%%%%%%%%%%%%%%%%%%%%%%%%%%%
\end{document}